# Parameterized Complexity Results for General Factors in Bipartite Graphs with an Application to Constraint Programming[*]

Gregory Gutin[†]  Eun Jung Kim[‡]  Arezou Soleimanfallah[†]

Stefan Szeider[§]  Anders Yeo[†]

## Abstract

The NP-hard general factor problem asks, given a graph and for each vertex a list of integers, whether the graph has a spanning subgraph where each vertex has a degree that belongs to its assigned list. The problem remains NP-hard even if the given graph is bipartite with partition $U \uplus V$, and each vertex in $U$ is assigned the list $\{1\}$; this subproblem appears in the context of constraint programming as the consistency problem for the extended global cardinality constraint.

We show that this subproblem is fixed-parameter tractable when parameterized by the size of the second partite set $V$. More generally, we show that the general factor problem for bipartite graphs, parameterized by $|V|$, is fixed-parameter tractable as long as all vertices in $U$ are assigned lists of length 1, but becomes W[1]-hard if vertices in $U$ are assigned lists of length at most 2. We establish fixed-parameter tractability by reducing the problem instance to a bounded number of acyclic instances, each of which can be solved in polynomial time by dynamic programming.

## 1 Introduction

To find in a given graph a spanning subgraph (or *factor*) that satisfies certain degree constraints is a fundamental task in combinatorics that entails several classical polynomial-time solvable problems such as PERFECT MATCHING (the factor is 1-regular), $r$-FACTOR (the factor is $r$-regular), and $(a, b)$-FACTOR (the degree of each vertex $v$ in the factor lies in a given interval $(a_v, b_v)$). Lovász [9, 10] introduced the following NP-hard problem which generalizes all mentioned factor problems:

GENERAL FACTOR

*Instance:* A graph $G = (V, E)$ and a mapping $K$ that assigns to each vertex $v \in V$ a set $K(v) \subseteq \{0, \ldots, d(v)\}$ of integers.

*Question:* Is there a subset $F \subseteq E$ such that for each vertex $v \in V$ the number of edges in $F$ incident with $v$ is an element of $K(v)$?

The problem remains NP-hard even for bipartite graphs $G = (U \uplus V, E)$ where $K(u) = \{1\}$ for all $u \in U$ and $K(v) = \{0, 3\}$ for all $v \in V$. Cornuéjols [5] obtained a dichotomy result that classifies

---

the complexity of all GENERAL FACTOR problems that are formed by restricting the sets $K(v)$ to a fixed class $\mathcal{C}$ of sets of integers. For each class $\mathcal{C}$ the corresponding problem is either polynomial or NP-complete.

In this paper we study the parameterized complexity of GENERAL FACTOR for bipartite graphs $G = (U \uplus V, E)$ parameterized by the size of $V$. Our main results can be summarized as follows.

> The problem GENERAL FACTOR for bipartite graphs $G = (U \uplus V, E)$, parameterized by the size of $V$, is
>
> (1) fixed-parameter tractable if $|K(u)| \leq 1$ for all $u \in U$;
> (2) W[1]-hard if $|K(u)| \leq 2$ for all $u \in U$.

We establish result (1) by a novel combination of concepts from polynomial-time algorithmics (alternating cycles) with concepts from fixed-parameter algorithmics (data reduction and annotation).

Next we briefly discuss an application of our fixed-parameter tractability result. Constraint Programming (CP) is a general-purpose framework for combinatorial problems that can be solved by assigning values to variables such that certain restrictions on the combination of values are satisfied; the restrictions are formulated by a combination of so-called *global constraints*. For example the global constraint ALLDIFFERENT enforces that certain variables must all be assigned to mutually different values. The Catalog of Global Constraints [1] lists hundreds of global constraints that are used to model various real-world problems. For several global constraints the consistency problem (i.e., deciding whether there exists an allowed value assignment) is NP-complete [3]. It is an interesting line of research to study such global constraints under the framework of parameterized complexity. We think that global constraints are an excellent platform for deploying efficient fixed-parameter algorithms for real-world applications.

An important global constraint is the *extended global cardinality constraint* (or *EGC constraint*, for short), also known as *global_cardinality* [1], *egcc* [3], *distribution* [4], and *card_var_gcc* [14]. Let $X$ be a finite set of variables, each variable $x \in X$ given with a finite set $D(x)$ of possible values. An EGC constraint over $X$ is specified by a mapping that assigns to each value $d \in D := \bigcup_{x \in X} D(x)$ a set $K(d)$ of non-negative integers. The constraint is consistent (or satisfiable) if one can assign each variable $x \in X$ a value $\alpha(x) \in D(x)$ such that $|\alpha^{-1}(d)| \in K(d)$ holds for all values $d \in D$. The consistency problem for EGC constraints can clearly be expressed as an instance $(G, K')$ of GENERAL FACTOR where $G$, the *value graph* of the constraint [18], is the bipartite graph $(X \uplus D, \{ xd : d \in D(x) \})$ and $K'$ is the degree list assignment defined by $K'(x) = \{1\}$ for all $x \in X$ and $K'(d) = K(d)$ for all $d \in D$ (see Figure 1 for an example). Hence our result (1) renders the consistency problem for EGC constraints fixed-parameter tractable when parameterized by the number $|D|$ of values.

## 1.1 Related Work

The parameterized complexity of EGC constraints was first studied by Samer and Szeider [15] using the treewidth of the value graph as the parameter. For value graphs of bounded degree it is easy to see that the consistency problem is fixed-parameter tractable for this parameter, as one can express the restrictions imposed by the sets $K(v)$ in monadic second-order logic, and use Courcelle's Theorem. However, for graphs of unbounded degree the problem is W[1]-hard. That instances of unbounded degree but bounded treewidth are solvable in non-uniform polynomial time (i.e., the consistency problem is in XP) can be shown by means of an extension of Courcelle's Theorem [16]. A further parameterization of GENERAL FACTOR was considered by Mathieson and Szeider [11], taking as the parameter the number of edges that need to be deleted to obtain the general factor. The problem is W[1]-hard in general but fixed-parameter tractable for graphs of bounded degree.

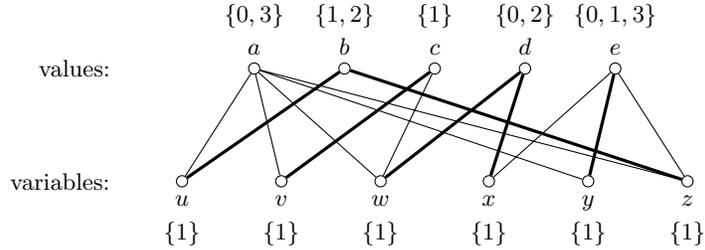

Figure 1: Value graph of an EGC constraint with variables $u, \dots, z$ and values $a, \dots, e$. For instance, variable $u$ has the domain $D(u) = \{a, b\}$. The constraint is satisfied by the assignment $u = b$, $v = c$, $w = d$, $x = d$, $y = e$, and $z = b$ which corresponds to the general factor indicated by bold edges.

The parameterized complexity of other global constraints were studied by Bessiere et al. [2]. Whether global constraints admit polynomial kernels was the subject of recent studies [8, 17].

In the context of parameterized complexity it is interesting to mention the results of van Hoeve et al. [19] who compare various algorithms for the SEQUENCE constraint (a global constraint that is important for various scheduling problems). Although the consistency problem is polynomial for this constraint, it turns out that a fixed-parameter algorithm outperforms the polynomial-time algorithm on several realistic instances.

## 1.2 Notation and Preliminaries

Unless otherwise stated, all graphs considered are finite, simple, and undirected. We denote a graph $G$ with vertex set $V$ and edge set $E$ by $G = (V, E)$ and write $V(G) = V$ and $E(G) = E$. We denote an edge between two vertices $u$ and $v$ by $uv$ or equivalently $vu$. For a set $F$ of edges and a vertex $v$ we write $N_F(v) = \{ u : uv \in F \}$ and we write $d_F(v)$ for the number of edges in $F$ that are incident with $v$. For a graph $G$ we also write $N_G(v) = N_{E(G)}(v)$ and $d_G(v) = d_{E(G)}(v)$, and we omit the subscripts if the context allows.

A *degree list assignment* $K$ is a mapping that assigns to each vertex $v \in V(G)$ a set $K(v) \subseteq \{0, \dots, d_G(v)\}$. A set $F \subseteq E(G)$ is a *general $K$-factor* of $G$ if $d_F(v) \in K(v)$ holds for each $v \in V(G)$. Sometimes it is convenient to identify a set $F \subseteq E(G)$ with the spanning subgraph $(V(G), F)$ of $G$.

An instance of a *parameterized problem* $L$ is a pair $(I, k)$ where $I$ is the *main part* and $k$ is the *parameter*; the latter is usually a non-negative integer. $L$ is *fixed-parameter tractable* if there exist a computable function $f$ and a constant $c$ such that instances $(I, k)$ can be solved in time $O(f(k)n^c)$ where $n$ denotes the size of $I$. FPT is the class of all fixed-parameter tractable decision problems.

A *parameterized reduction* is a many-one reduction where the parameter for one problem maps into the parameter for the other. More specifically, problem $L$ reduces to problem $L'$ if there is a mapping $R$ from instances of $L$ to instances of $L'$ such that (i) $(I, k)$ is a YES-instance of $L$ if and only if $(I', k') = R(I, k)$ is a YES-instance of $L'$, (ii) $k' = g(k)$ for a computable function $g$, and (iii) $R$ can be computed in time $O(f(k)n^c)$ where $f$ is a computable function, $c$ is a constant, and $n$ is the size of $I$. The parameterized complexity classes $W[1] \subseteq W[2] \subseteq \cdots \subseteq XP$ are defined as the closure of certain parameterized problems under parameterized reductions. There is strong theoretical evidence that parameterized problems that are hard for classes $W[i]$ are not fixed-parameter tractable.

For more background on parameterized complexity we refer to other sources [6, 7, 12].

## 2 Fixed-Parameter Tractability

This section is devoted to the proof of our fixed-parameter tractability result. Let BIPARTITE GENERAL FACTOR WITH SINGLETONS denote the problem GENERAL FACTOR restricted to instances $(G, K)$ where $G = (U \uplus V, E)$ is bipartite and $|K(u)| \leq 1$ for all $u \in U$. We will show the following:

**Theorem 1.** BIPARTITE GENERAL FACTOR WITH SINGLETONS *parameterized by the size of $V$ is fixed parameter tractable.*

Let $(G, K)$ be an instance of BIPARTITE GENERAL FACTOR WITH SINGLETONS with $G = (U \uplus V, E)$ and $V = \{v_1, \ldots, v_k\}$. Clearly we may assume that $K(v) \notin \{\emptyset, \{0\}\}$ for all $v \in U \uplus V$: if $K(v) = \emptyset$ then $G$ has no general $K$-factor, and if $K(v) = \{0\}$ then we can delete $v$ from $G$. Thus, in particular for each $u \in U$ we have $K(u) \in \{\{1\}, \ldots, \{k\}\}$.

### 2.1 General Factors of Edge-Weighted Graphs

Key to our algorithm for BIPARTITE GENERAL FACTOR WITH SINGLETONS is the transformation to a more general "annotated" problem on edge-weighted graphs that allows a more succinct representation.

Let $G$ be a graph. A (positive integral) *edge-weighting* $\rho$ of $G$ is a mapping that assigns to each edge $e \in E(G)$ a non-negative integer $\rho(e)$. We refer to a pair $(G, \rho)$ as an *edge-weighted graph*. For a vertex $v$ of $G$ we define $d_\rho(v)$ as the sum of $\rho(e)$ over all edges incident with $v$ (or 0 if $v$ has no incident edges). As usual, $d_G(v)$ denotes the number of edges incident with $v$, the degree of $v$. Let $K$ be a degree list assignment of $G$. We define a general $K$-factor of an edge-weighted graph $(G, \rho)$ by using $\rho(e)$ as the "capacity" of an edge $e$. More precisely, we say that an edge-weighting $\varphi$ is a *general $K$-factor* of the edge-weighted graph $(G, \rho)$ if (i) $\varphi(e) \leq \rho(e)$ holds for all edges $e$ of $G$ and (ii) $d_\varphi(v) \in K(v)$ for all $v \in V(G)$. Evidently this definition generalizes the above definition of general $K$-factors for unweighted graphs (by considering an unweighted graph as an edge-weighted graph where each edge has weight 1, and a set $F$ of edges as an edge-weighting that assigns each edge in $F$ the weight 1, and all other edges the weight 0). By GENERAL FACTOR FOR EDGE-WEIGHTED GRAPHS we refer to the obvious generalization of the decision problem GENERAL FACTOR to edge-weighted graphs.

In the following we will present several *reduction rules* that take as input an instance $I = (G, \rho, K)$ of GENERAL FACTOR FOR EDGE-WEIGHTED GRAPHS and produce as output an instance $I' = (G', \rho', K')$ of the same problem (or rejects $I$ as a no-instance). We say that a reduction rule is *sound* if it always holds that either both $I$ and $I'$ are no-instances or both are yes-instances (or in case of rejection, $I$ is indeed a no-instance). A reduction rule is *polynomial* if we can decide in polynomial time whether it applies to $I$ and we can compute $I'$ in polynomial time if the rule applies.

### 2.2 Contractions of Modules

Let $(G, \rho, K)$ be an instance of GENERAL FACTOR FOR EDGE-WEIGHTED GRAPHS. For an integer $c \geq 1$ we call a subset $M \subseteq V(G)$ a *c-module* if

1. $M$ is nonempty and independent;

2. $K(v) = \{c\}$ for all $v \in M$;

3. all vertices in $M$ have exactly the same neighbors;

4. $\rho(e) = 1$ holds for all edges $e \in E(G)$ with one end in $M$.

**Reduction Rule 1.** *Let $M$ be a $c$-module of $(G, \rho, K)$. Obtain a new instance $(G', \rho', K')$ by replacing $M$ with a new vertex $u_M$ that is adjacent with the same vertices as the vertices in $M$. Set $\rho'(e) = |M|$ for all edges $e$ incident with $u_M$ and $\rho'(e') = \rho(e')$ for all other edges $e'$. Set $K'(u_M) = \{c|M|\}$ and $K'(v) = K(v)$ for all other vertices.*

**Lemma 1.** *Reduction Rule 1 is sound and polynomial.*

*Proof.* Let $\varphi$ be a general $K$-factor of $(G, \rho)$. We define $\varphi'(u_M w) = \sum_{v \in M} \varphi(vw)$ for edges $u_M w$ that are incident with $u_M$ and $\varphi'(e) = \varphi(e)$ for all other edges. Observe that $\varphi'(u_M w) \leq |M| = \rho'(u_M w)$ and $d_{\varphi'}(u_M) = \sum_{v \in M} d_\varphi(v) = c|M| \in K'(u_M)$, hence $\varphi'$ is a general $K'$-factor of $(G', \rho')$.

Conversely, let $\varphi'$ be a general $K'$-factor of $(G', \rho')$. Let $M = \{u_1, \ldots, u_s\}$ and let $N = \{v_1, \ldots, v_t\}$ be the set of neighbors of $u_M$.

We define an edge-weighting $\varphi$ of $G$. For $0 \leq i \leq t$ let $S_i = \sum_{i'=1}^{i} \varphi'(u_M v_{i'})$; thus $S_0 = 0$ and $S_t = cs$. For $1 \leq i \leq t$ and $1 \leq j \leq s$ we set

$$\varphi(v_i u_j) = \begin{cases} 1 & \text{if } j \equiv S_{i-1} + l \pmod{s} \text{ for some } 1 \leq l \leq \varphi'(u_M v_i); \\ 0 & \text{otherwise.} \end{cases}$$

For $e \in E(G) \cap E(G')$ we set $\varphi(e) = \varphi'(e)$. Since $\varphi'(u_M v_i) \leq s$ for $1 \leq i \leq t$ this definition is correct. To see that $\varphi$ is a general $K$-factor of $G$ we note that $d_\varphi(v_i) = d_{\varphi'}(v_i)$ for all $1 \leq i \leq t$, and $d_\varphi(u_j) = d_{\varphi'}(u_M)/s = c \in K(u_j)$ for all $1 \leq j \leq s$.

As it is obvious that the rule is polynomial, the lemma follows. □

### 2.3 Acyclic General Factors

Let $\varphi$ be a general $K$-factor of an edge-weighted graph $(G, \rho)$. We say that an edge $e \in E(G)$ is *full* in $\varphi$ if $\varphi(e) = \rho(e)$ and $\varphi(e) > 0$, an edge $e \in E(G)$ is *empty* in $\varphi$ if $\varphi(e) = 0$.

The *skeleton* of $\varphi$ is the spanning subgraph $G_\varphi$ of $G$ with $E(G_\varphi) = \{\, e \in E(G) : 0 < \varphi(e) < \rho(e) \,\}$; i.e., $E(G_\varphi)$ consists of all edges that are neither full nor empty. The *full skeleton* of $\varphi$ is the spanning subgraph $G_\varphi^+$ of $G$ with $E(G_\varphi^+) = \{\, e \in E(G) : 0 < \varphi(e) \leq \rho(e) \,\}$; i.e., $E(G_\varphi^+)$ consists of all edges that are not empty. We say that $\varphi$ is *acyclic* if its skeleton $G_\varphi$ contains no cycles (i.e., is a forest), $\varphi$ is *fully acyclic* if its full skeleton $G_\varphi^+$ contains no cycles.

**Lemma 2.** *If a bipartite edge-weighted graph $(G, \rho)$ has a general $K$-factor, then it also has an acyclic general $K$-factor.*

*Proof.* Let $G = (U \uplus V, E)$ where $U = \{u_1, \ldots, u_p\}$ and $V = \{v_1, \ldots, v_k\}$. For an edge-weighting $\varphi$ of $G$ and a pair $v_i, u_j$ with $v_i u_j \notin E(G)$ we define $\varphi(v_i u_j) = 0$. With each edge-weighting $\varphi$ of $G$ we associate a vector $A(\varphi)$ defined as follows:

$$\begin{aligned} A(\varphi) = (&\varphi(v_1 u_1), \varphi(v_1 u_2), \ldots, \varphi(v_1 u_p), \\ &\varphi(v_2 u_1), \varphi(v_2 u_2), \ldots, \varphi(v_2 u_p), \\ &\ldots \\ &\varphi(v_k u_1), \varphi(v_k u_2), \ldots, \varphi(v_k u_p)). \end{aligned}$$

Let $\varphi$ be a general $K$-factor of $G$ such that $A(\varphi)$ is lexicographically maximal among all vectors of general $K$-factors of $G$. (A vector $(a_1, \ldots, a_n)$ is lexicographically larger than $(b_1, \ldots, b_n)$ if for some $i$, $a_i > b_i$ and $a_j = b_j$ for all $j < i$.) We are going to show that $\varphi$ is acyclic.

We will use the following idea. Suppose the skeleton $G_\varphi$ contains a cycle. Then we alternately increase and decrease the weights of the edges on the cycle by one. Since the graph is bipartite, the cycle is of even length, and so every vertex on the cycle is incident with an edge of increased weight and an edge with decreased weight. Thus the weighted degree of each vertex remains the same. By changing the weights of the edges on the cycle we obtain a new general $K$-factor whose associated vector is lexicographically larger than the vector associated with $\varphi$, a contradiction to the choice of $\varphi$.

Suppose to the contrary that the skeleton $G_\varphi$ contains a cycle $C = v_{j_1} u_{i_1} \ldots v_{j_t} u_{i_t} v_{j_1}$. Without loss of generality, we may assume that $j_2 = \min\{j_1, \ldots, j_t\}$. Moreover we may assume that $i_1 < i_2$ since otherwise we can consider the reverse of $C$ instead.

We define a new general $K$-factor $\varphi'$ of $(G, \rho)$ by setting $\varphi'(v_{j_l} u_{i_{l-1}}) = \varphi(v_{j_l} u_{i_{l-1}}) + 1$ and $\varphi'(v_{j_l} u_{i_l}) = \varphi(v_{j_l} u_{i_l}) - 1$ for $1 \leq l \leq t$ (computing indices modulo $t$), and $\varphi'(e) = \varphi(e)$ for all other edges. That $\varphi'$ is indeed a general $K$-factor follows from the following observations:

(i) $d_{\varphi'}(w) \in K(w)$ holds for all $w \in U \uplus V$ since $d_{\varphi'}(w) = d_\varphi(w) \in K(w)$;

(ii) For each $e \in E(C)$ we have $\varphi'(e) \leq \varphi(e) + 1 \leq \rho(e)$ since $e \in E(C) \subseteq E(G_\varphi)$ and therefore $e$ is not full in $\varphi$.

(iii) For each $e \in E(C)$ we have $0 \leq \varphi(e) - 1 \leq \varphi'(e)$ since $e \in E(C) \subseteq E(G_\varphi)$ and therefore $e$ is not empty in $\varphi$.

Furthermore, we observe that $\varphi'(v_{j_2} u_{i_1}) > \varphi(v_{j_2} u_{i_1})$, but all entries in the vector $A(\varphi')$ before $\varphi(v_{j_2} u_{i_1})$ remain the same as in $A(\varphi)$. Hence $A(\varphi')$ is lexicographically larger than $A(\varphi)$, a contradiction to our assumption that $A(\varphi)$ is lexicographically maximal. This proves that $G_\varphi$ is indeed acyclic. □

Let $I = (G, \rho, K)$ be an instance of GENERAL FACTOR FOR EDGE-WEIGHTED GRAPHS and $X \subseteq E(G)$. For a vertex $v$ of $G$ let $d_X(v)$ be the sum of $\rho(e)$ over all edges $e \in X$ that are incident with $v$. Let $I - X$ denote the instance $(G^X, \rho^X, K^X)$ obtained from $I$ by deleting $X$, decreasing the capacities $\rho$ of all edges not in $X$ by one, and updating the degree list assignment assuming the edges in $X$ are full. More precisely, we set

$$
\begin{aligned}
G^X &= G - X, \\
\rho^X(e) &= \max(\rho(e) - 1, 0) \quad \text{for all } e \in E(G^X), \\
K^X(v) &= \{c - d_X(v) : c \in K(v),\ c - d_X(v) \geq 0\} \quad \text{for all } v \in V(G^X).
\end{aligned}
$$

**Lemma 3.** *Let $I = (G, \rho, K)$ be an instance of* GENERAL FACTOR FOR EDGE-WEIGHTED GRAPHS, *$X \subseteq E(G)$, and $I - X = (G^X, \rho^X, K^X)$. Then the following two statements are equivalent.*

1. *$(G, \rho)$ has an acyclic general $K$-factor $\varphi$ such that $X$ is precisely the set of full edges of $\varphi$.*

2. *$(G^X, \rho^X)$ has a fully acyclic general $K^X$-factor.*

*Proof.* Let $\varphi$ be an acyclic general $K$-factor of $(G, \rho)$ such that $X$ is precisely the set of full edges of $\varphi$. Thus, the skeleton $G_\varphi$ contains no cycles. The restriction $\varphi'$ of $\varphi$ to $G^X$ is clearly a general $K^X$-factor of $(G^X, \rho^X)$. Moreover, since $(G^X)^+_{\varphi'} = G_\varphi$, it follows that $\varphi'$ is fully acyclic.

Conversely, let $\varphi'$ be a fully acyclic general $K^X$-factor of $(G^X, \rho^X)$. Let $\varphi$ be the edge-weighting of $(G, \rho)$ defined by

$$
\begin{aligned}
\varphi(e) &= \varphi'(e) \quad \text{for } e \in E(G^X), \text{ and} \\
\varphi(e) &= \rho(e) \quad \text{for } e \in X.
\end{aligned}
$$

Clearly $\varphi$ is a general $K$-factor of $(G, \rho)$ where all edges in $X$ are full. Since the capacities of edges in $G^X$ were decreased by one, no edge of $G$ outside $X$ can be full with respect to $\varphi$. Hence $X$ is precisely the set of full edges of $\varphi$. As above, we have $(G^X)^+_{\varphi'} = G_\varphi$, and so $\varphi$ is acyclic. □

**Lemma 4.** *Let $(G, \rho)$ be a bipartite edge-weighted graph and $K$ a degree list assignment such that for each edge $uv$ of $G$ we have $K(u) = \{\rho(uv)\}$ or $K(v) = \{\rho(uv)\}$. If $(G, \rho)$ has a general $K$-factor, then it also has a fully acyclic general $K$-factor.*

*Proof.* Assume that $(G, \rho)$ has a general $K$-factor. By Lemma 2, $(G, \rho)$ has an acyclic general $K$-factor $\varphi$. Let $X$ be the set of full edges of $\varphi$. Let $I - X = (G^X, \rho^X, K^X)$ and $\varphi^X$ the restriction of $\varphi$ to $G^X$. By the proof of Lemma 3, $\varphi^X$ is a general $K^X$-factor of $(G^X, \rho^X)$ which is fully acyclic. For each edge $uv \in X$ we have $K^X(u) = \{0\}$ or $K^X(v) = \{0\}$, hence at least one of the ends of any $uv \in X$ is of degree 1 in the full skeleton $G_\varphi^+$. Since $G_\varphi^+$ can be obtained by adding the edges in $X$ to the forest $G_\varphi$, it follows that also $G_\varphi^+$ is a forest, i.e., $\varphi$ is fully acyclic. □

### 2.4 Eliminating Vertices of Low Degree

**Reduction Rule 2.** *Assume that $G$ has a vertex $v$ of degree 0. If $0 \notin K(v)$, then reject the instance; if $0 \in K(v)$ then delete $v$ from $G$ and let $G' = G - v$, $\rho' = \rho$, and $K'$ the restrictions of $K$ to $V(G')$.*

**Reduction Rule 3.** *Assume $G$ has a vertex $v$ of degree 1. Let $u$ be the neighbor of $v$. We let $G' = G - v$, $\rho'$ the restriction of $\rho$ to $E(G')$, and we put*

$$K'(u) = \{\, c_u - c_v : c_u \in K(u),\ c_v \in K(v),\ c_v \leq \min(\rho(uv), c_u) \,\}$$

*and $K'(w) = K(w)$ for all $w \in V(G') \setminus \{u\}$.*

The proof of the following lemma is obvious.

**Lemma 5.** *Reduction Rules 2 and 3 are sound and polynomial.*

**Proposition 1.** GENERAL FACTOR FOR EDGE-WEIGHTED GRAPHS *can be solved in polynomial time for edge-weighted forests.*

*Proof.* Let $I = (G, \rho, K)$ be an instance of GENERAL FACTOR FOR EDGE-WEIGHTED GRAPHS such that $G$ is a forest. If $V(G) \neq \emptyset$, $G$ has a vertex of degree $\leq 1$, and hence Reduction Rule 2 or 3 applies, and we obtain in polynomial time an equivalent instance with one vertex less which is again a forest (or we reject the instance). By at most $|V(G)|$ applications of the rules we either reject the instance ($I$ is a no-instance) or we eliminate all the vertices ($I$ is a yes-instance). Thus the result follows by repeated application of Lemma 5. □

### 2.5 The Algorithm

It remains to put together the above results to show that BIPARTITE GENERAL FACTOR WITH SINGLETONS parameterized by the size of $V$ is fixed-parameter tractable.

Let $(G, K)$ with $V(G) = U \uplus V$ be the given instance of the problem. As explained above we can consider $(G, K)$ as an instance $I = (G, \rho, K)$ of GENERAL FACTOR FOR EDGE-WEIGHTED GRAPHS letting $\rho(e) = 1$ for all $e \in E(G)$. Let $|V| = k$.

1. We partition $U$ into maximal sets $U_1, \ldots, U_p$ such that each $U_i$ is a $c$-module for some $1 \leq c \leq k$.

2. We apply Reduction Rule 1 with respect to the modules $U_1, \ldots, U_p$ and obtain an instance $I_M = (G_M, \rho_M, K_M)$ of GENERAL FACTOR FOR EDGE-WEIGHTED GRAPHS where $G_M = (U_M \uplus V, E_M)$ is a bipartite graph with $p + k$ vertices.

3. We guess a set $X \subseteq E_M$ of edges and consider the instance $I_M^X = I_M - X = (G_M^X, \rho_M^X, K_M^X)$.

4. We guess a spanning forest $T$ of $G_M^X$ and consider the instance $I^{(T)} = (T, \rho^{(T)}, K^{(T)})$ where $\rho^{(T)}$ is the restriction of $\rho_M^X$ to $T$ and $K^{(T)} = K_M^X$. We check if $(T, \rho^{(T)})$ has a general $K^{(T)}$-factor using Proposition 1 (i.e., applying the Reduction Rules 2 and 3).

    If $(T, \rho^{(T)})$ has general $K^{(T)}$-factor, then we stop and output YES.

5. If none of the guesses for $X$ and $T$ produces the answer YES, we stop and output NO.

**Theorem 2.** *Given a bipartite graph $G = (U \uplus V, E)$, $k = |V|$, $n = |E|$, and a degree list assignment $K$ with $|K(u)| \leq 1$ for all $u \in U$, we can decide whether $G$ has a general $K$-factor in time $2^{k^2 2^k + k^2}(k+1)^{k 2^k + k} \cdot n^{O(1)}$.*

*Thus, BIPARTITE GENERAL FACTOR WITH SINGLETONS parameterized by the size of $V$ is fixed-parameter tractable.*

*Proof.* We show that the above algorithm decides correctly and in the claimed time bound whether $G$ has a general $K$-factor.

Consider the following sequence of statements.

(1) $(G, \rho)$ has a general $K$-factor.

(2) $(G_M, \rho_M)$ has a general $K_M$-factor.

(3) $(G_M, \rho_M)$ has an acyclic general $K_M$-factor.

(4) There is some $X \subseteq E(G_M)$ such that $(G_M^X, \rho_M^X)$ has a fully acyclic general $K_M$-factor.

(5) There is some $X \subseteq E(G_M)$ and some spanning forest $T$ of $G_M^X$ such that $T$ has a general $K_M^X$-factor.

By previous results all five statements are equivalent. In particular, (1) $\Leftrightarrow$ (2) follows by Lemma 1, (2) $\Leftrightarrow$ (3) follows by Lemma 2, and (3) $\Leftrightarrow$ (4) follows by Lemma 3. The equivalence (4) $\Leftrightarrow$ (5) is obvious. The correctness of the algorithm thus follows from Proposition 1, it remains to bound its running time.

We may assume that $U$ contains no isolated vertices as such vertices can be ignored without changing the problem. Each $U_i$ is defined by some degree constraint $c \in \{1, \ldots, k\}$ and a nonempty subset of $V$, hence $p \leq k(2^k - 1)$. Since $G_M$ has at most $k^2 2^k$ edges, there are at most $2^{k^2 2^k}$ possible choices for $X$. In a spanning forest $T$, each vertex has at most one parent. Each vertex in $U$ has $k+1$ alternatives for its parent (including not having one), and each vertex in $V$ has at most $k 2^k$ alternatives for its parent (including not having one). Thus, each $G_M^X$ has at most $O((k+1)^{k 2^k}(k 2^k)^k)$ possible spanning forests $T$. It follows that we apply Proposition 1 at most $O(2^{k^2 2^k}(k+1)^{k 2^k}(k 2^k)^k) = O(2^{k^2 2^k + k^2}(k+1)^{k 2^k + k})$ times. □

**Corollary 1.** *Given a bipartite graph $G = (U \uplus V, E)$, $k = |V|$, $n = |E|$, and a degree list assignment $K$ with $K(u) = \{1\}$ for all $u \in U$, we can decide in time $2^{k^2}(k+1)^{k 2^k + k} \cdot n^{O(1)}$ whether $G$ has a general $K$-factor.*

*Proof.* We use a simplified version of the above algorithm. In view of Lemma 4 we do not need to guess a set $X$ of full edges in order to be able to restrict our scope to fully acyclic general $K$-factors. Thus we may skip step 3 of the algorithm and save a factor of $2^{k^2 2^k}$ in the running time. □

# 3 W[1]-Hardness

This section is devoted to establishing the W[1]-hardness result. Let BIPARTITE GENERAL FACTOR WITH PAIRS denote the problem GENERAL FACTOR restricted to instances $(G, K)$ where $G = (U \uplus V, E)$ is bipartite and $K(u) \in \{\, \{0, r\} : 1 \leq r \leq |V| \,\}$ holds for all $u \in U$. We will show the following:

**Theorem 3.** BIPARTITE GENERAL FACTOR WITH PAIRS *parameterized by the size of $V$ is W[1]-hard.*

We give a parameterized reduction from the following problem (also called MULTICOLORED CLIQUE) which is known to be W[1]-complete [13] for parameter $k$.

> PARTITIONED CLIQUE
> *Instance:* A $k$-partite graph $G = (V_1 \uplus \ldots \uplus V_k, E)$ where $|V_i| = n$ for all $1 \leq i \leq k$.
> *Parameter:* The integer $k$.
> *Question:* Is there a $k$-clique (a complete subgraph on $k$ vertices) in $G$?

For this reduction we need to ensure that exactly one vertex $v_i$ is selected from each partite set $V_i$, $1 \leq i \leq k$, and that $v_i$ and $v_j$ are adjacent for all $1 \leq i < j \leq k$. For the first requirement we shall use the following gadget construction.

Given a set $A$ of non-negative integers with $M = \max(A)$ and a number $r > 0$, we construct a complete bipartite graph $G_{A,r} = (U' \uplus V', E')$ with $U' = \{u_1, \ldots, u_M\}$, $V' = \{v_0, v_1, \ldots, v_r\}$, and a degree list assignment $K$, setting $K(u_i) = \{0, r+1\}$ for $1 \leq i \leq M$ and $K(v_0) = A$. We do not impose any degree restrictions on the vertices $v_1, \ldots, v_r$, hence we put $K(v_j) = \{0, \ldots, M\}$ for $1 \leq j \leq r$. We call the graph $G_{A,r}$ together with the degree list assignment $K$ a *selection gadget*, and we refer to the vertices $v_1, \ldots, v_r$ as the *outputs* of the gadget.

**Lemma 6.** *If a set $F$ of edges forms a general $K$-factor of a selection gadget $G_{A,r}$ then all outputs are incident to the same number $\alpha$ of edges in $F$, and $\alpha \in A$. Conversely, for each $\alpha \in A$ there exists a general $K$-factor $F$ of $G_{A,r}$ such that each output is incident with exactly $\alpha$ edges of $F$.*

*Proof.* Suppose that $F$ is a general $K$-factor of $G_{A,r} = (U' \uplus V', E)$. Let $U'' = \{\, u_i : d_F(u_i) = r+1, 1 \leq i \leq M \,\}$ and $d_F(v_0) = \alpha \in A$. Clearly $|U''| = \alpha$. Hence $d_F(v_j) = |U''| = \alpha$ holds for all $1 \leq j \leq r$.

Conversely, let $\alpha \in A$. Then $F = \{\, u_i v_j : 1 \leq i \leq \alpha, 0 \leq j \leq r \,\}$ forms a general $K$-factor of $G_{A,r}$ with $d_F(v_j) = \alpha$ for all outputs $v_j$. □

Let $A$ be a set of non-negative integers, $N = \max(A) + 1$, $A' = \{\, N\alpha : \alpha \in A \,\}$ and $r, r' \geq 0$ two numbers. We take two vertex-disjoint selection gadgets $G_{A,r+1}$ and $G_{A',r'+1}$ and identify one output $v$ of the first with one output $v'$ of the second gadget. Let us call this identified vertex $q$. We define $K(q) = \{\, a + Na : a \in A \,\}$. We call this new gadget a *double selection gadget* $G_{A,r,r'}$. We consider the outputs of $G_{A,r+1}$ and $G_{A',r'+1}$ except $v$ and $v'$ as the outputs of $G_{A,r,r'}$. We call the $r$ outputs that originate from $G_{A,r+1}$ the *lower outputs*, and the $r'$ outputs that originate from $G_{A',r'+1}$ the *upper outputs* of $G_{A,r,r'}$. If $U \uplus V$ denotes the vertex set of $G_{A,r,r'}$, then $|V| = r + r' + 3$.

**Lemma 7.** *If a set $F$ of edges is a general $K$-factor of a double selection gadget then all lower outputs are incident to the same number $\alpha$ of edges in $F$, all upper outputs are incident to the same number $\beta$ of edges in $F$, and we have $\alpha \in A$ and $\beta = \alpha N$. Conversely, for each $\alpha \in A$ there is a general $K$-factor $F$ such that all lower outputs are incident to $\alpha$ edges in $F$, and all upper outputs are incident to $\alpha N$ edges in $F$.*

*Proof.* Let $G_{A,r,r'}$ be a double selection gadget constructed from two selection gadgets $G_{A,r+1}$ and $G_{A',r'+1}$, and let $F$ be a general factor of $G_{A,r,r'}$. Let $\alpha = |N_F(q) \cap V(G_{A,r+1})|$ and $\beta = |N_F(q) \cap V(G_{A',r'+1})|$. Clearly $\alpha + \beta = d_F(q) \in K(q)$. By Lemma 6 we have $\alpha \in A$, $\beta \in A'$, $d_F(v) = \alpha$ for all lower outputs $v$ and $d_F(v') = \beta$ for all upper outputs $v'$, thus the first part of the lemma is shown. The second part follows easily by using the second part of Lemma 6 twice. □

Next we describe the parameterized reduction from PARTITIONED CLIQUE to BIPARTITE GENERAL FACTOR WITH PAIRS that uses the double selection gadgets. Let $G = (V_1 \uplus \ldots \uplus V_k, E)$ be an instance of PARTITIONED CLIQUE, and assume $n = |V_i|$ for $1 \leq i \leq k$. We write $V_i = \{v_1^i, \ldots, v_n^i\}$. For every $1 \leq i \leq k$, we take a copy $H_i$ of the double selection gadget $G_{A,r,r'}$ where $A = \{1, \ldots, n\}$, $r = i - 1$ and $r' = k - i$. For each pair $1 \leq i < j \leq k$ we identify an upper output of $H_i$ and a lower output of $H_j$. We denote the identified vertex as $h_{i,j}$. We can choose the identified pairs in such a way that finally each output is identified with exactly one other output. Let $H = (U_H \uplus V_H, E_H)$ be the bipartite graph constructed in this way. We define a degree list assignment $K$ where each identified vertex $h_{i,j}$, $1 \leq i < j \leq k$, gets assigned the list $\{N\alpha + \beta : v_\alpha^i v_\beta^j \in E(G), \alpha, \beta \in \{1, \ldots, n\}\}$, and all other vertices inherit the list assigned to them in the definition of a double selection gadget. Thus $(H, K)$ is an instance of GENERAL FACTOR that satisfies the properties as stated in Theorem 3 (in fact, for all $u \in U_H$ we have $K(u) \in \{\{0, r\} : 2 \leq r \leq k + 1\}$). Furthermore, we have $|V_H| = \binom{k}{2} + 3k$ as $V_H$ contains $\binom{k}{2}$ identified vertices and each $H_i$, $1 \leq i \leq k$, contributes 3 more vertices to $V_H$. Therefore the new parameter $k' = |V_H|$ of the BIPARTITE GENERAL FACTOR WITH PAIRS instance is indeed a function of the old parameter $k$ of the PARTITIONED CLIQUE instance. Furthermore, it is easy to check that $|U_H| = k \cdot 2n(n+2)$ and clearly $(H, K)$ can be obtained from $G$ in polynomial time. It remains to show that the reduction is correct:

**Lemma 8.** *$H$ has a general $K$-factor if and only if $G$ has a $k$-clique.*

*Proof.* Let $F$ be a general $K$-factor of $H$. For $1 \leq i \leq k$ let $F_i = F \cap E(H_i)$ and observe that $F_i$ is a general factor of $H_i$. Thus, by the first part of Lemma 7 there is some $a_i \in A = \{1, \ldots, n\}$ such that $d_{F_i}(v) = a_i$ for each lower output $v$ of $H_i$ and $d_{F_i}(v') = Na_i$ for each upper output $v'$ of $H_i$. Let $1 \leq i < j \leq k$ and consider the identified vertex $h_{i,j}$. We have $d_F(h_{i,j}) = Na_i + a_j$. Since $K(h_{i,j}) = \{N\alpha + \beta : v_\alpha^i v_\beta^j \in E(G)\}$, it follows that $v_{a_i}^i v_{a_j}^j \in E(G)$. Hence $C = \{v_{a_i}^i : 1 \leq i \leq k\}$ induces a clique in $G$.

Conversely, assume that $C \subseteq V(G)$ induces a $k$-clique in $G$. Since $G$ is $k$-partite, $C$ contains exactly one vertex $v_{x_i}^i$ from each set $V_i$, $1 \leq i \leq k$. By the second part of Lemma 7, each $H_i$, $1 \leq i \leq k$, has a general factor $F_i$ such that $d_{F_i}(v) = x_i$ for each lower output $v$ and $d_{F_i}(v') = Nx_i$ for each upper output $v'$. Let $F = \bigcup_{i=1}^{k} F_i$. Since for each pair $1 \leq i < j \leq k$ we have $v_{x_i} v_{x_j} \in E(G)$, it follows that $d_F(h_{i,j}) = x_j + Nx_i \in K(h_{i,j})$, hence $F$ is indeed a general $K$-factor of $H$. □

With Lemma 8 we have shown that our reduction is correct, thus Theorem 3 is established.

## 4 Conclusion

We have studied the parameterized complexity of general factor problems for bipartite graphs $G = (U \uplus V, E)$ where the size of the sets $K(u)$ for $u \in U$ is bounded by a small constant and where $|V|$ is the parameter. There are various further variants of general factor problems whose parameterized complexities would be interesting to explore, for example, one could consider $|U|$ instead of $|V|$ as the parameter. A further possibility is to restrict $K(v)$ for all vertices $v$ of one or both partite sets to a fixed class $\mathcal{C}$ of sets of integers, similar to Cornuejols's dichotomy result [5]. It would be interesting to reveal fixed-parameter tractable general factor problems that are W[1]-hard without the restriction of $K(v)$ to a fixed class $\mathcal{C}$ and NP-hard without the parameterization.

**Acknowledgements** Research of Gutin, Kim and Yeo was supported in part by an EPSRC grant. Research of Gutin was supported in part by the IST Programme of the European Community, under the PASCAL 2 Network of Excellence. Research of Soleimanfallah was supported in part by the Overseas Research Students Award Scheme (ORSAS). Research of Szeider was supported by the European Research Council, grant reference 239962.